**An Ising model for the thermal and dynamic properties of supercooled liquids and the glass transition**

**Ralph V. Chamberlin [1],*** 

[1]    Department of Physics, Arizona State University, Tempe, AZ  85287-1504 USA; chamberl@asu.edu

*    Correspondence: chamberl@asu.edu

**Abstract:** We describe the behavior of an Ising model with orthogonal dynamics, where changes in energy and changes in alignment never occur during the same Monte Carlo (MC) step. This orthogonal Ising model (OIM) allows conservation of energy and conservation of (angular) momentum to proceed independently, on their own preferred time scales. The OIM also includes a third type of MC step that makes or breaks the interaction between neighboring spins, facilitating an equilibrium distribution of bond energies. MC simulations of the OIM mimic more than twenty distinctive characteristics that are commonly found above and below the glass temperature, $T_g$. Examples include a specific heat that has hysteresis around $T_g$, out-of-phase (loss) response that exhibits primary ($\alpha$) and secondary ($\beta$) peaks, super-Arrhenius $T$ dependence for the $\alpha$-response time ($\tau_\alpha$), and fragilities that increase with increasing system size ($N$). Mean-field theory for energy fluctuations in the OIM yields a critical temperature ($T_c$) and a novel expression for the super-Arrhenius divergence as $T \to T_c$: $\ln(\tau_\alpha) \sim 1/(1 - T_c/T)^2$. Because this divergence is reminiscent of the Vogel-Fulcher-Tammann (VFT) law squared, we call it the "VFT2 law." A modified Stickel plot, which linearizes the VFT2 law, shows that at high $T$ where mean-field theory should apply, only the VFT2 law gives qualitatively consistent agreement with measurements of $\tau_\alpha$ (from the literature) on five glass-forming liquids. Such agreement with the OIM suggests that several basic features govern supercooled liquids. The freezing of a liquid into a glass involves an underlying 2$^{nd}$-order transition that is broadened by finite-size effects. The VFT2 law for $\tau_\alpha$ comes from energy fluctuations that enhance the pathways through an entropy bottleneck, not activation over an energy barrier. Values of $\tau_\alpha$ vary exponentially with inverse $N$, consistent with the distribution of relaxation times deduced from measurements of $\alpha$ response. System sizes found via the $T$ dependence of $\tau_\alpha$ from simulations and measurements are similar to sizes of independently relaxing regions (IRR) measured by nuclear magnetic resonance (NMR) for simple-molecule glass-forming liquids. The OIM elucidates the key ingredients needed to interpret the thermal and dynamic properties of amorphous materials, while providing a broad foundation for more-detailed models of liquid-glass behavior.

## 1. Introduction

Our goal is to find the simplest model that mimics the greatest number of features commonly shown by supercooled liquids [1-3]. The simplest microscopic picture for a thermodynamic phase transition is the homogeneous Ising model on an infinite lattice [4,5]. However, heterogeneity is known to be a crucial characteristic of supercooled liquids [6-14], and it is still an open question as to whether freezing of a liquid into a glass involves an underlying transition. Furthermore, in a strict sense, the Ising model applies only to binary degrees of freedom ("spins"), such as uniaxial magnetic or electric dipoles, binary alloys, or lattice gases; not molecules that may move continuously in all directions. Nevertheless, the Ising model can be a useful starting point for studies in statistical physics [15,16], similar to the ideal gas for thermodynamics or the fruit fly for genetics. Therefore, we seek the minimal modifications to the Ising model that allow it to match the broadest range of features in the liquid-glass transformation. Here, we extend the standard Ising model by adding finite-size effects, a thermal distribution of interaction energies, and orthogonal constraints on the dynamics. We show that this "orthogonal Ising model" (OIM) mimics more than twenty features commonly found in supercooled liquids and the glass transition, and we discuss key insights gained from the model.

The concept of simulating Ising models with kinetic constraints was introduced by Kawasaki [17]. The Kawasaki exchange step can be thought of as exchanging the location of two neighboring spins without changing their alignments, thus ensuring constant net alignment during each step. In its original application this constraint was used to conserve both types of particles in a binary mixture, yielding particle diffusion near a critical point. Subsequent studies of kinetic Ising models using other constraints exhibit various characteristics of supercooled liquids [18-21], including super-Arrhenius activation of the primary ($\alpha$) response time ($\tau_\alpha$) as a function of temperature ($T$), and stretched-exponential relaxation of the alignment as a function of time ($t$). Another approach for adapting the Ising model to supercooled liquids involves mean-field theory on a thermal distribution of finite systems [22,23]. Results from this mesoscopic mean-field theory include peak response frequencies ($f_p$) that follow the Vogel-Fulcher-Tammann (VFT) law for super-Arrhenius activation, $\alpha$-response peaks that are asymmetrical due to relaxation rates that vary exponentially with inverse size, and an underlying phase transition that is smeared out by finite-size effects. Here we combine finite-size effects with a novel set of constraints on Ising dynamics to yield a simplified, but microscopic model that mimics more than twenty distinctive properties of the liquid-glass transition. Although the constraints we use for OIM can be thought of as empirical assumptions, we describe evidence indicating that they are justified by fundamental physics.

Kawasaki exchange is the standard constraint that we add to MC simulations, and one way to create transition rates with a dynamical hierarchy that allows realistic simulations of physical processes using MC [24,25]. A less-common modification involves MC steps that change the alignment, without changing the net energy. When combined, these two constraints yield "orthogonal dynamics," where steps that change energy never change the net alignment, and vice versa. Note that orthogonal dynamics does not prevent correlations between changes in alignment and energy, as they may have a preference to occur on nearby steps. A recent investigation studied an Ising model with orthogonal dynamics, plus a local entropy bath that makes states with the same energy effectively indistinguishable [26]. That model, which utilized a one-dimensional (1-D) chain of spins to facilitate analytic calculations, matches several details in the thermal noise found in qubits. Here, we first extend the orthogonal dynamics to a 3-D Ising model, as needed for an underlying transition at a critical temperature $T_c > 0$. Then we remove the local entropy bath, consistent with local energy states that are distinguishable, as expected for the variety of distinct environments in amorphous systems. Finally, we let the Ising model have intermittent interactions between spins, allowing an equilibrium distribution of interaction energies. We call this model – having orthogonal dynamics, finite-size effects, and intermittent interactions – the orthogonal Ising model, OIM.

The remainder of this paper is separated into six sections. In section 2, we present the foundations of the OIM. In section 3 we apply mean-field theory to obtain analytic expressions for behavior expected from the OIM. In section 4 we describe details for simulating the OIM. In section 5 we present results showing that the OIM exhibits many common features found in the thermal and dynamic response of most glass-forming liquids. In section 6, we discuss these results and their basic implications for the distinctive behavior associated with the liquid-glass transition, as well as compare and contrast the OIM to other models and interpretations. Finally, in section 7, we give a brief review.

## 2. The Orthogonal Ising Model, Foundations

*2.1 Standard Ising model*

We start with the standard Ising model for binary spins on a simple-cubic lattice [27]. We use cube-shaped systems, with sides of length $\ell$ and periodic boundary conditions, yielding a total of $N = \ell^3$ spins with $3N$ bonds between spins. Each spin may be aligned up ($\sigma_i = +1$) or down ($\sigma_i = -1$). The total energy is

$$E = -\frac{1}{2}\sum_{i=1}^{N} H_i \sigma_i \qquad (1)$$

Here, the sum is over all $N$ spins in the system ($\sigma_i$), with the factor of ½ needed to remove double counting. Using $J_{ij}$ as the interaction energy between $\sigma_i$ and $\sigma_j$, the local field is the sum over all 6 nearest neighbors:

$$H_i = \sum_{j=1}^{6} J_{ij} \sigma_j \qquad (2)$$

The magnetic moment (total alignment) is

$$M = \sum_{i=1}^{N} \sigma_i \qquad (3)$$

In the standard Ising model a uniform exchange interaction is used for all bonds, $J_{ij} = J$. One modification for the OIM is to allow a Boltzmann weighted MC step where $J_{ij}$ may go to 0 if initially $J_{ij} = J$, or $J_{ij}$ may go to $J$ if initially $J_{ij} = 0$. These intermittent interactions add entropy to the system and facilitate an equilibrium distribution of interaction energies [26].

Standard MC simulations of the Ising model [28] start by choosing a spin at random from the lattice $\sigma_i$, then calculating the change in energy ($\Delta E_i = 2H_i\sigma_i$) to flip the spin ($\sigma_i \to -\sigma_i$). The spin flip is accepted only if the Metropolis criterion is met: $e^{-\Delta E_i/kT} > [0,1)$, where $[0,1)$ is a uniformly-distributed random number between 0 and 1. Usually this procedure is repeated for $N$ steps to yield one MC sweep (MCS). Then the simulation is repeated for $Q$ MCS, until average values from the system reach their equilibrium values to within some desired accuracy. Two such averages are $\bar{E} = \sum_{q=1}^{Q} E_q/Q$ and $\bar{M} = \sum_{q=1}^{Q} M_q/Q$, with subscript $q$ referring to values averaged over the $q^{th}$ MCS. Thermodynamic limits of these values can be found by simulating systems of increasing size, then extrapolating $N \to \infty$. On a simple-cubic lattice of effectively infinite size, this standard Ising model has a phase transition at a Curie temperature of $T_C \approx 4.5115J/k$, which is lower than the mean-field Weiss temperature of $6J/k$ found by extrapolating from high-$T$ behavior. Because the OIM transition is smeared out by heterogeneity and finite-size effects, we also extrapolate high-$T$ behavior to define the critical temperature of the OIM, $T_c$ (note lower-case $c$). In fact, due to finite-size effects, fluctuations, and intermittent interactions, this $T_c$ is always lower than the Curie temperature, $T_c < T_C$, sometimes by as much as an order of magnitude.

At most temperatures, simple simulations of the standard Ising model relax exponentially towards equilibrium. Near $T_C$, however, critical slowing down often yields a power-law relaxation as a function of time. Such slowing is usually considered a nuisance, and cluster algorithms have been developed to accelerate the approach to equilibrium near $T_C$. In cases where slow relaxation is of interest, kinetic Ising models are often used. Some types of kinetic constraints exhibit well-known glass-like behavior, including stretched-exponential relaxation and super-Arrhenius activation [18-21]. Often the slow response is studied via time-dependent relaxation from non-equilibrium initial states. For our studies, we usually simulate the OIM for long enough times to reach thermal equilibrium, then we analyze the time-dependent fluctuations using the fluctuation-dissipation theorem to obtain frequency-dependent behavior in equilibrium.

*2.2 Orthogonal dynamics, with intermittent interactions*

To extend the usefulness of the standard Ising model we add three modifications: finite-size effects from nanoscale regions, orthogonal dynamics from distinct conservation laws, and intermittent bonds from a thermal distribution of interaction energies. These modifications are motivated by practical considerations, empirical evidence, and fundamental physics. Specifically, finite-size systems are technically required for computer simulations, empirically expected for dynamic heterogeneity [6-14], and theoretically justified by thermal equilibrium in the nanocanonical ensemble [26]. Orthogonal dynamics allows separate time scales for two basic laws of classical mechanics that govern most systems: conservation of momentum and conservation of energy. The orthogonality is implemented by requiring that each MC step may change either the spin alignment, or the spin energy, but never both. Although the resulting separation of time scales is consistent with their fundamental limit – spin flips involve electromagnetic interactions mediated by virtual photons while heat flow involves phonons – other factors usually control actual time scales. For example, in nuclear magnetic resonance (NMR), alignment changes are governed by precession rates due to local fields, while heat flow is governed by (usually much slower) spin-lattice relaxation. Similarly, dielectric response measurements on supercooled liquids show time scales for dipole rotation that span the entire spectrum – from alpha ($\alpha$) response that may be slower than 1 s, through intermediate beta ($\beta$) response, to microscopic processes that are faster than 1 ns – while energy equilibration is usually dominated by the $\alpha$ response. Indeed, nonlinear dielectric response measurements show that the time scale for dipole rotation and energy equilibration can differ by more than an order of magnitude [29-32]. Other experiments show that glass-forming liquids have a separation of time scales between linear and rotational motion [33,34], suggesting that the two laws conserving momentum may also be uncoupled. Even in idealized single crystals, molecular dynamics simulations show that energy is persistently localized by anharmonic interactions [35], implying that energy localization will be even stronger in disordered systems. In any case, orthogonal dynamics in a simulation does not prohibit correlations between energy change and dipole rotation, it simply separates them so that they may proceed independently on their own preferred time scales.

One justification for non-interacting spins is if they are highly localized to distinct sites, halting the exchange interaction. For molecules, a related mechanism comes from the correlated dynamics needed for various interactions. For example, a van der Waals-like interaction requires in-phase fluctuations of induced dipoles, which is easily achieved for two molecules that are close enough to avoid any time delay that would yield a retarded van der Waals interaction, and isolated enough to limit incoherent thermal fluctuations. However, it is unlikely that all molecules in a sample can simultaneously have coherent fluctuations, especially if there is an intervening thermal bath. Likewise, the London dispersion force that yields realistic van der Waals-like interactions requires quantum effects that involve overlapping wave functions between interacting molecules, and again this coherence is broken by wavefunctions that are localized. Classically, molecular dynamics simulations have shown that, even in idealized single crystals, anharmonic interactions yield significant deviations from standard fluctuation relations due to energy localization [35], which will be much stronger in non-crystalline systems.

The OIM is based on the same equations for energy and alignment as the standard Ising model, Eqs. (1)-(3). Although we focus on the explicit finite-size effects in the OIM, finite-size systems are unavoidable in computer simulations. Thus, the most novel features of the OIM are its orthogonal dynamics and intermittent interactions. The orthogonality is constructed by ensuring that energy changes and spin flips never occur during the same MC step. Intermittent interactions arise by assuming a third type of MC step that may change spins from interacting to non-interacting, or vice versa. OIM dynamics starts by choosing a spin at random from the lattice, $\sigma_i$, then proceeds with one of three options. To maximally mix all three types of steps, each option is chosen at random with probability of 1/3.

The first option is to attempt a spin flip, $\sigma_i \rightarrow -\sigma_i$. The spin flip succeeds only if the local field at the site is zero, $H_i = 0$, so that the energy will not change. This requires that there be an even number of interacting neighbors, with half the interacting neighbors up and the other half down. Note that this may occur even in fully-aligned systems if $J_{ij} = 0$ for all neighbors. The second option is to attempt a Kawasaki spin exchange between $\sigma_i$ and one neighboring spin, $\sigma_j$. Without bias, this $\sigma_j$ is chosen at random from the three nearest-neighbor spins along the positive axes. If $\sigma_i$ and $\sigma_j$ are aligned, any exchange is trivial, with no change in alignment or energy. If $\sigma_i$ and $\sigma_j$ are anti-aligned, an exchange is attempted only if the spins are interacting, consistent with $J_{ij} \neq 0$ due to exchange. The spin exchange is accepted only if the total change in energy ($\Delta E_{ij} = 2(H_i\sigma_i + H_j\sigma_j + 2)$) meets the Metropolis criterion, $e^{-\Delta E_{ij}/kT} > [0,1)$. The third option is an attempt to change energy by changing the bond between $\sigma_i$ and $\sigma_j$. Again, this energy change is accepted only if the Metropolis criterion is met, $e^{\pm J\sigma_i\sigma_j/kT} > [0,1)$, with the + (–) sign chosen if initially the spins are interacting (non-interacting). An additional constraint is to accept bond changes only if the net alignment around $\sigma_i$ is zero, $\sum_{j=1}^{6}\sigma_j = 0$, limiting bond changes to regions of high entropy. This constraint causes the irreversibility below the hysteresis temperature ($T_h$), without substantively altering the behavior above $T_h$. Similar to Kawasaki exchange, changing the bond changes the net energy, but never the alignment. When averaged over sufficiently long times, this third option yields an equilibrium distribution of interaction energies between spins. Figure 1 shows a 2-D version of this OIM.

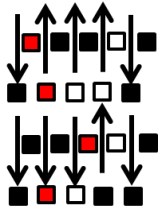

**Figure 1.** Sketch of a 2-D version of the orthogonal Ising model with intermittent interactions and periodic boundary conditions. Binary spins may be up or down, as shown by the arrows. Intermittent interactions may be low energy (black), high energy (red), or no energy (white). For the configuration shown, isoenergetic spin flips can occur only for the three middle spins on the bottom row. Specifically, the first of these three spins may flip because it has two low-energy and two high-energy interactions, while the other two spins have one low-energy and one high-energy interaction. Kawasaki exchange may occur between any pair of interacting spins (connected by a black or red square) whenever the energy change meets the Metropolis criterion. As a third option, an interaction may end by changing red or black to white, or begin by changing white to red or black.

Like many simulations of the standard Ising model, we utilize a simple-cubic lattice inside a cube-shaped system having sides of length $\ell$. Thus, the system has a total of $N = \ell^3$ spins, but $< 3N$ interacting bonds between the spins. Unlike most simulations of the standard Ising model, we focus on finite-size effects from the dependence on $N$.

## 3. The Orthogonal Ising Model, Simulations

*3.1 Simulation details*

Each simulation of the OIM is made at a run temperature ($T_r$) within the range $0.4 < kT_r/J < 30$. (Recall that the Curie temperature for the standard Ising model on a simple-cubic lattice of infinite size is $kT_C/J \approx$ 4.5115.) At the starting temperature $T_0$, to thoroughly mix the spin states before the first simulation run is begun, the system is initialized for $10^5$ MCS using the standard Metropolis algorithm, without local constraints. Subsequent temperatures are usually decreased by a constant factor, $T_1=aT_0$, where $0.84 \leq a \leq 0.98$. Most simulations are made in a set of 10 temperatures $T_0$, $T_1$ ... $T_9$. However, for studying hysteresis, some simulations are made using similar steps down from $T_0$ through $T_g$, followed by steps up through $T_g$ to $T_0$ utilizing the constant factor $1/a$, so that simulations down and up occur at the same $T$.

Each simulation run proceeds for time $Q = \tau \times 10^P$ MCS, where $\tau = 2^{17} = 131{,}072$ is a multiple of two to optimize the fast-Fourier transform. Here, the power of ten ($P$) is an integer that yields an "integration time" ($10^P$) that is fixed for a given set of simulations, then may be changed for subsequent simulations to cover a wide range of response times from fast ($P = 0$) to slow ($P = 4$-$6$). The maximum value of $P$ is limited by the size of the system and the available computer time. Time-dependent quantities are recorded after each integration time in a moving average, averaged over the preceding $10^P$ MCS. The main quantities we study are the energy per spin $\varepsilon = E/N$, alignment per spin $m = M/N$, and fraction of interacting bonds $b = \sum_{i=1}^{N} H_i/6NJ$. Each simulation run yields $\tau$ sets of these quantities, with moving averages that render time-dependent behavior over long times while maintaining a manageable number of data points. An average value of each quantity is found by averaging all of its moving averages at a given temperature.

*3.2 Numerical analysis of simulations*

From the per-spin alignment values ($m_t$), averaged over the preceding $10^P$ MCS to give the value at $t$, we use standard techniques to obtain the out-of-phase susceptibility (loss) as a function of frequency, $\chi''(f)$. First, a power-spectral density is found from the magnitude squared of a discrete Fourier transform:

$$S(f) = \frac{1}{\tau^2} \left| \sum_{t=0}^{\tau-1} m_t e^{-\frac{2\pi i f t}{\tau}} \right|^2 \qquad (4)$$

Because $S(f)$ can be quite noisy, we use a smoothing procedure that involves linear regression applied to $S(f)$ on a log-log scale. We start with a set of frequencies that have the same frequency range as the Fourier transform ($\log(1) = 0$ to $\log(\tau/2) = 4.816487$), but are chosen to be evenly spaced on a logarithmic scale, e.g. $\log(f_0) = 0$, $\log(f_1) \approx 0.00732$, $\log(f_2) \approx 0.01452$ .... For each of these frequencies, $f_\delta$, all data points within $\log(f_\delta) \pm 0.2$ are fit with a linear function, then evaluated at $f_\delta$ to yield a smoothed set of value, $S(f_\delta)$. Next, the frequency-dependent loss is found using the fluctuation-dissipation theorem, $\chi''(f) = \chi_0 f S(f_\delta)/kT$, with $\chi_0$ an amplitude factor. Note that $\chi''(f)$ is presented only for equilibrium fluctuations above $T_h$, where the fluctuation-dissipation theorem remains valid.

Finally, loss spectra from different integration times at each temperature are put onto a common scale by adjusting their magnitudes and frequencies. Specifically, if $\chi_P''(f_P)$ is the loss spectrum from a simulation having an integration time of $10^P$, frequencies are shifted to the same scale using $f = f_P/10^P$. Similarly, magnitudes of fluctuations that are averaged over $10^P$ sweeps vary as $1/\sqrt{10^P}$, so that the magnitude squared (e.g. power-spectral density) varies as $1/10^P$. Together these results can be written as $\chi(f) = 10^P \chi_P(f_P/10^P)$. Merging loss spectra from different integration times is facilitated by finding a common set of frequencies. Again, we use frequencies that are evenly-spaced on a logarithmic scale, but now over a coarser grain, e.g. $\log(f_0) = 0.00$, $\log(f_1) = 0.05$, $\log(f_2) = 0.10$ .... Interpolation is used to find the value of loss from each spectrum at all common frequencies encompassed by the spectrum. Combining spectra is done with a Gaussian weighting factor involving the logarithm of frequency, $w = e^{-[\log(f) - \log(f_{mid})]^2}$. Thus, $w = 1$ at $f = f_{mid}$ (the mid-point frequency of each spectrum on a logarithmic scale) where $S(f)$ is usually best defined, falling to $w \approx 1/330$ at the lowest and highest frequencies, where $|\log(f) - \log(f_{mid})| = \log(\tau/2)/2 \approx 2.408$. Note that the sharpness of $w$ can be altered by changing the width of the Gaussian, but we find similar results for a wide range of widths, indicating that this detail is not important. The inverse of this $w$ squared is treated as a sample variance, so that contributions to a spectrum at frequency $f_s$ can be written as:

$$\chi(f_s) = \frac{10^0 \chi_0(f_s) e^{-2[\log(f_s)-\log(f_{mid0})]^2} + 10^1 \chi_1(f_s) e^{-2[\log(f_s)-\log(f_{mid1})]^2} + \cdots}{e^{-2[\log(f_s)-\log(f_{mid0})]^2} + e^{-2[\log(f_s)-\log(f_{mid1})]^2} + \cdots} \quad (5)$$

There is sufficient overlap between spectra that the loss at most frequencies comes from averaging 3-5 independent values from different integration times. The weighting factor and merging that yield Eq. (5) allow several spectra to be smoothly melded into a single spectrum that can cover more than 10 orders of magnitude in frequency, and is consistent with the original spectra.

## 4. The Orthogonal Ising Model, Theory

*4.1 Primary response from energy fluctuations in mesoscopic mean-field theory*

We use mean-field theory of energy fluctuations in a finite-size system containing $N$ spins [22,23] to derive theoretical expressions for the $T$ dependence of the characteristic time for the $\alpha$ response, $\tau_\alpha$. Note that in mean-field theory, because all fluctuations become negligible if $N \to \infty$, finite-size effects are required for dynamics. In fact, most real systems have independently relaxing regions (IRR) inside bulk samples with length scales of 1-3 nanometers [12], yielding $n \ll 1000$ particles. (Lower-case $n$ is used for IRR inside bulk samples.) From experiments, $\tau_\alpha$ is found by inverting the peak-loss frequency, $\tau_\alpha = 1/f_p$, with $f_p$ found from the peak dielectric loss. (A factor of $2\pi$ that would simply shift the results on a logarithmic scale is neglected.) From simulations, $f_p$ is deduced from the Fourier transform of the time-dependent equilibrium fluctuations in spin alignment, also yielding $\tau_\alpha = 1/f_p$.

We attribute $\alpha$ response to alignment inversions that change the sign of the magnetization, e.g $m_t < 0$ to $m_{t+1} > 0$, or vice versa. Orthogonal dynamics requires that spin flips never change the energy. Thus, the sequence of spin flips that invert the alignment yielding $\alpha$ response must never directly involve energy activation. However, spin flips occur only if there is an equal number of up and down interacting neighbors, so that inversions of $m$ usually coincide with increases in energy. Indeed, we find significant correlations between energy increases and $\alpha$ response, but various features indicate that the mechanism involves energy fluctuations that facilitate passing through an entropy bottleneck, not activation over an energy barrier. Such entropy bottlenecks and barriers have long been studied for the dynamics of complex systems [36-39].

Standard fluctuation theory [40,41] treats the probability of finding a change in entropy $\Delta S$, yielding a rate for fluctuations $R \propto e^{\Delta S/k}$. To calculate $R$, we expand the change in entropy as a function of energy to second order: $\Delta S = \frac{\partial S}{\partial E}(E - \bar{E}) + \frac{1}{2}\frac{\partial^2 S}{\partial E^2}(E - \bar{E})^2$. Standard thermodynamic relations give $\frac{\partial S}{\partial E} = \frac{1}{T}$ and $\frac{\partial^2 S}{\partial E^2} = -\frac{1}{T^2 C_V}$, where $C_V = \frac{\partial \bar{E}}{\partial T}$ is the heat capacity. The linear term in the expansion of $\Delta S$ is balanced by a linear change of energy $(\bar{E} - E)$ in the thermal reservoir, yielding Boltzmann's factor $e^{-(\bar{E}-E)/kT}$ that is accommodated by Metropolis weighting in the simulations. Here we focus on the quadratic term that comes from finite-size fluctuations with no analogue in an infinite reservoir. In general, spin flips are most likely to occur if energy is near to zero, $E \to 0$, where the net local field is most likely to be zero. Hence, replacing $\Delta S$ by $\Delta S_0 = -\frac{1}{2}\frac{1}{T^2 C_V}(\bar{E})^2$, we obtain an expression for a peak $\alpha$-response time of:

$$\tau_\alpha \propto e^{\bar{E}^2/(2kT^2 C_V)} \quad (6)$$

To calculate $\bar{E}$ and $C_V$ for the exponent of Eq. (6) we use mesoscopic mean-field theory [22,23], a Landau-like theory for phase transitions of finite-size systems. To quantify finite-size effects we start with the free energy *per spin*, $f(m)$. This $f(m)$ includes the mean-field interaction energy per spin, $-6Jm^2/2$, combined with the binomial coefficient for the degeneracy of these energies. Using Stirling's formula for the factorials to quartic order in $m$, and $T_c = 6J/k$ as the mean-field critical temperature, the alignment-dependent contributions to free-energy per particle can be written as:

$$\frac{f(m)}{kT} \sim \frac{1}{2}\ln(1-m^2) + \frac{m}{2}\ln\left(\frac{1+m}{1-m}\right) - \frac{T_c}{2T}m^2 \approx \frac{1}{2}\left(1 - \frac{T_c}{T}\right)m^2 + \frac{1}{12}m^4$$

The average energy is found from the thermally-weighted integral of $m^2$ divided by the partition function

$$\bar{E} = -\frac{NJ}{2}\left[\frac{\int_0^1 m^2 \exp[-Nf(m)/kT]dm}{\int_0^1 \exp[-Nf(m)/kT]dm}\right] \approx -\frac{NJ}{2}\left[\frac{\int_0^\infty x^{1/2}\exp[-Nf(x)/kT]dx}{\int_0^\infty x^{-1/2}\exp[-Nf(x)/kT]dx}\right]$$

Here, the approximation on the right comes from making a change of variables to $x = m^2$, and extending the upper limits on the integrals to $\infty$. These integrals can be evaluated in terms of special functions (integral 3.462 in [42]) by writing the argument in the exponents in the form:

$$\frac{Nf(x)}{kT} = \frac{N}{2}\left(1 - \frac{T_c}{T}\right)x + \frac{N}{12}x^2 = \gamma x + \beta x^2$$

where $\gamma = (N/2)(1 - T_c/T)$ and $\beta = N/12$. Then, using Eq. 19.3.7 from [43], in terms of parabolic cylinder functions, $U(a,z) = D_{-a-1/2}(z)$ where $z = \gamma/\sqrt{2\beta} = \sqrt{3N/2}(1 - T_c/T)$, the average energy can be written as:

$$\bar{E} \approx -\frac{NJ}{2}\frac{1}{2\sqrt{2\beta}}\left[\frac{U(1,z)}{U(0,z)}\right]$$

At high temperatures ($T > T_c$) if the system is not too small ($N>10$), $z > 1$ favors an asymptotic expansion for the parabolic cylinder function. Using Eq. 19.8.1 in [43] $U(a,z) \sim e^{-z^2/4}z^{-a-1/2}\{1 - (a+1/2)(a+3/2)/2z^2 + \cdots\}$, yields: $\bar{E} \approx -\frac{NJ}{2}\sqrt{\frac{3}{2N}}\left[\frac{\left(1 - \frac{15}{8z^2} + \frac{945}{128z^4}\right)}{z\left(1 - \frac{3}{8z^2} + \frac{105}{128z^4}\right)}\right] \approx -\frac{J}{2}\frac{1}{\left(1 - \frac{T_c}{T}\right)}\left[1 - \frac{3}{2z^2} + \frac{6}{z^4}\right]$, or:

$$\bar{E} \approx -\frac{J}{2}\frac{1}{\left(1 - \frac{T_c}{T}\right)}\left[1 - \frac{1}{N\left(1 - \frac{T_c}{T}\right)^2} + \frac{8}{3N^2\left(1 - \frac{T_c}{T}\right)^4}\right] \qquad (7)$$

Note that to lowest order, the total energy is intensive, independent of *N*, a consequence of mean-field theory above the transition where contributions to energy come only from finite-size fluctuations. Furthermore, if this lowest-order term was utilized as an activation energy in an Arrhenius law, $\tau_\alpha \propto e^{-\bar{E}/kT}$ yields the VFT law. However, we find that $\alpha$ response is due to energy fluctuations that allow the system to traverse through an entropy bottleneck, not over a barrier. From Eq. (7), the heat capacity is:

$$C_V = \frac{\partial \bar{E}}{\partial T} \approx \frac{k}{2}\frac{(T_c/T)^2}{(1 - T_c/T)^2}\left[1 - \frac{3}{N\left(1 - \frac{T_c}{T}\right)^2} + \frac{40}{3N^2\left(1 - \frac{T_c}{T}\right)^4}\right]$$

Thus, the characteristic $\alpha$-response time (inverse of relaxation rate) can be written as:

$$\tau_\alpha \propto \exp\left[\frac{k(\bar{E})^2}{2(kT)^2 C_v}\right] \approx \exp\left\{\frac{[1 - 1/[N(1 - T_c/T)^2]]^2}{4[1 - 3/[N(1 - T_c/T)^2]]}\right\} \approx \exp\left\{\frac{1}{4}\left[1 + \frac{1}{N(1 - T_c/T)^2} - \frac{4}{N^2(1 - T_c/T)^4}\right]\right\}. \qquad (8)$$

From the 1/*N*-dependent term in Eq. (8) we define a curvature coefficient $C = 4N$. Then, using $\tau_\infty$ as the *α*-response time of an infinite region, $N \to \infty$, the *α*-response time of finite regions can be written as:

$$\tau_\alpha = \tau_\infty \exp\left[\frac{1/C}{(1 - T_c/T)^2}\right] \tag{9}$$

Note that increasing *C* increases the curvature on an Angell plot, hence *C* increases monotonically with increasing fragility. Empirically, when allowed to be an adjustable parameter we find: $C/N \ll 4$ from simulations and $C/n \lesssim C/N$ from experiments. Mechanisms that could cause *C* to be smaller than predicted by this simplified theory involve *T* dependences that may amplify the influence of *C*. One example is that $T_c$ increases with decreasing *T* due to the increasing fraction of interacting bonds. Another example is the assumption yielding Eq. (6) that $E \to 0$, whereas from simulations we find that *α* response proceeds at lower energies ($E < 0$) that are *T* dependent. For experiments there is an additional *T* dependence in *n* [12] that may further reduce *C*. Nevertheless, the dominant *T* dependence of Eq. (9) is a tendency to diverge as $\ln[\tau_\alpha] \sim 1/(1 - T_c/T)^2$. Although similar to various generalized VFT formulas [44,45], to our knowledge the specific *T* dependence of Eq. (9) has not previously been proposed. As examples, the Bässler formula [46] predicts $\ln[\tau_\alpha] \sim 1/T^2$ without a finite critical temperature, while the VFT law predicts a linear divergence $\ln[\tau_\alpha] \sim (1/T)/(1 - \theta/T)$, where $\theta$ is the Vogel temperature. Because Eq. (9) diverges exponentially with the inverse of reduced temperature (offset from $T_c$) squared, we call it the "VFT2 law." Similarly, if the $4/[N^2(1 - T_c/T)^4]$ term from Eq. (8) is added to Eq. (9), we call it the "VFT4 law."

As a function of system size, *C* in Eq. (9) should be proportional to *N*. Theoretically, this *N* dependence is indicative of relaxation governed by fluctuations that decrease with increasing *N*, not activation over a barrier where heights increase with increasing *N*. Historically, this *N* dependence was found empirically for relaxation in random magnetic systems [47], and single-crystal ferromagnets [48,49]. Subsequent application to glass-forming liquids yielded an interpretation of the glass transition as an abrupt change in the size distribution [50]. Furthermore, the size distribution yields a distribution of response times that is significantly better than the Cole-Davidson or stretched-exponential formulas [51] for characterizing measured spectra. Here, we use this *N* dependence to deduce the size of IRR (*n*) inside bulk materials.

The prefactor in Eq. (9) can alter the *T* dependence of response times, especially in simulations at high *T*. In general, the rate $1/\tau_\infty$ comes from microscopic spin flips involving distinct local environments. At low *T* for simulations, and for measurements at all *T*, the dominant *T* dependence of $\tau_\alpha$ comes from the VFT2 divergence of Eq. (9), which will be the main focus of our analysis.

*4.2 Analysis of data*

A useful way to distinguish between formulas for $\tau_\alpha$ is to take *T*-dependent differentials, which eliminate an adjustable parameter and linearize the formulas. One such analysis that linearizes the VFT law, introduced by Stickel et al. [52-54], is to plot the square root of $[\Delta \ln(\tau_\alpha)/\Delta(1/T)]^{-1}$ as a function of $1/T$. However, most measurements show changes in slope on this Stickel plot, requiring multiple linear fits to encompass the entire range of data. Furthermore, detailed analyses on dozens of substances [55-57] indicate that measured behavior often deviates significantly from the VFT law. Equation (9) predicts a novel *T* dependence for $\tau_\alpha$. Indeed, Eq. (9) implies that the *cube* root of $[\partial \ln(\tau_\alpha)/\partial(1/T)]^{-1}$ is needed to linearize the response time as a function of $1/T$, which for finite differentials can be written as:

$$\left[\frac{\Delta \ln(\tau_\alpha)}{\Delta(T_c/T)}\right]^{-1/3} = \sqrt[3]{C/2}\left(1 - \frac{T_c}{T}\right) \tag{10}$$

When Eq. (10) is plotted as a function of $T_c/T$, both the intercept and magnitude of slope (|slope|) should be $\sqrt[3]{C/2} = (2N)^{1/3}$. Here we show that Eqs. (9) and (10) give good agreement with the measured $T$ dependence of $\tau_\alpha$ from several substances, and with simulations of the OIM.

Another consequence of deducing $\tau_\alpha$ from $T$-dependent fluctuations is that $C$ in Eqs. (9) and (10) is proportional to $N$. Thus, linear-response measurements or simulations of $\tau_\alpha$ as a function of $T$ provide information about the size of IRR. Furthermore, if the $T$ dependence of $\tau_\alpha$ is known for two (or more) sizes, extrapolation and/or interpolation can be used to quantitatively predict the size of IRR in other systems.

## 5. Results

### 5.1 Fluctuations as a function of time

Figure 2 shows time-dependent fluctuations in the alignment ($m$, black), energy ($\varepsilon/J$, red), and fraction of interacting spins ($b$, blue). The time scale in (A) is over an entire simulation run of 1.3M (Mega) MCS, and in (B) over 1.3k MCS centered around each alignment inversion. Simulations come from a system of $N = 64$ spins at $kT/J = 1.77$, about 60% below the Curie temperature of the standard Ising model on an infinite and homogeneous lattice. First note the two distinct behaviors shown by $m$ in Fig. 2 (A): fast fluctuations near one saturated alignment, with rare but abrupt inversions to the other alignment. These two types of behavior yield, respectively, the secondary ($\beta$) response at higher $f$, and the primary ($\alpha$) response at lower $f$, consistent with NMR measurements showing excess small-angle motions at higher $f$ [58]. Now, focus on some details of the behavior in (A) as $t \to 1.2$M MCS. Specifically, $m$ fluctuates around the down alignment for a relatively long time, then inverts to up at $t \lesssim 1.2$M MCS. The solid symbols in Fig. 2 (B) show this inversion in greater detail. Note the coincident behavior in $\varepsilon/J$ (red circles) and $m$ (black squares): first $\varepsilon/J$ fluctuates up as $m$ starts to increase, next $\varepsilon/J$ stays high as $m$ inverts from down to up, then $\varepsilon/J$ rises again each time $m$ attempts (without success) to invert back down. Although this coincidence seems to imply that alignment inversion requires energy activation over a barrier, orthogonal dynamics is constructed so that spin flips never involve energy activation. Therefore, we must examine the behavior more closely.

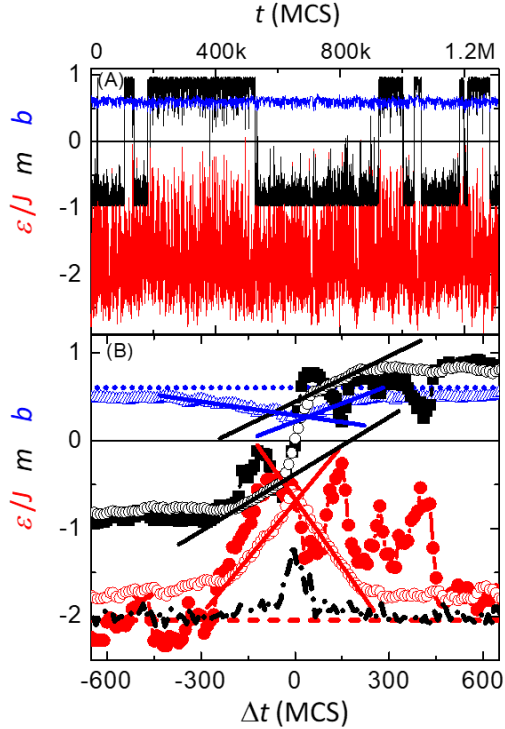

**Figure 2.** Time dependence of net alignment per spin (black), energy per spin (red), and fraction of interacting bonds (blue) from MC simulations of a system containing $N = 64$ spins at a temperature of $kT/J = 1.77$. The simulations utilized an integration time of $10^1$ MCS yielding (A) a total run time of 1.3M MCS, while (B) shows an expanded scale of 1.3k MCS around each alignment inversion. Open symbols in (B) show the average of 31 alignment inversions, with $\Delta t = 0$ defined by where $m$ changes sign from down to up, with up-to-down inversions inverted to add constructively to the average. Solid symbols in (B) are from the single down-to-up inversion at $t \lesssim 1.2$M MCS in (A). Note three unsuccessful attempts to revert back to down immediately after the successful inversion. Broken lines in (B) show: $\bar{\varepsilon}/J$ (dashed red), $\bar{m}$ (dotted blue), and $\delta\bar{m}/\delta t$ (dash-dotted black). Solid lines in (B) show linear fits to the open symbols over a range of times before, and after the inversion. Note that in (B), changes from the equilibrium of $b$ and $\delta\bar{m}/\delta t$ are multiplied by a factor of 5 for visibility.

Open symbols in Fig. 2 (B) show the average of 31 alignment inversions, from the simulation shown in Fig. 2 (A) and from a similar simulation at the same $T$. In Fig. 2 (B), $\Delta t = 0$ is defined by where the alignment changes sign from $m_t < 0$ to $m_{t+1} > 0$, with up-to-down inversions inverted to add constructively to the average. Again, because $\bar{\varepsilon}/J$ rises smoothly to a peak before $\bar{m}$ inverts sharply, it appears as though alignment inversion may involve energy activation; but spin flips must never change the energy, so the mechanism is more-subtle. First note that the rate at which the inversion occurs ($\delta\bar{m}/\delta t$, dash-dot black line) has a peak with width (FWHM) of less than 1/3 the FWHM for the peak in $\bar{\varepsilon}/J$ (open red circles), indicating that $m$ is not directly controlled by $\bar{\varepsilon}/J$. Next note the slopes of the solid lines in Fig. 2 (B), which come from linear fits to the open symbols over a comparable interval of times before and after the inversion. Qualitatively, even on the scale of Fig. 2 (B) it can be seen that the ratio in the slope before the peak divided by the magnitude of the slope after the peak, is <1 for fluctuations in energy, but >1 for alignment inversions. Quantitatively, this ratio in magnitudes is 0.85±0.03 for $\bar{\varepsilon}/J$ and 0.38±0.08 for $\bar{b}$, with 1.24±0.05 for $\bar{m}$. Thus, energy tends to fluctuate away from equilibrium slower than towards equilibrium, consistent with behavior governed by Boltzmann's factor, whereas alignment moves away from equilibrium faster than towards equilibrium, opposite to the behavior expected for activation over an energy barrier.

The $\alpha$ response in the OIM comes from net alignment passing through an entropy bottleneck, not activation over an energy barrier. In general, these two processes coincide because fluctuations up in energy enhance the likelihood of individual spins having an equal number of up- and down-interacting neighbors, thereby increasing the number of pathways through the bottleneck. This interpretation is consistent with the relative values of the slopes: $\bar{m}$ has a steep slope up as alignment inversion is increasingly accelerated when $\bar{\varepsilon}/J$ fluctuates slowly upwards, but $\bar{m}$ has a shallow slope down as it is increasingly retarded when $\bar{\varepsilon}/J$ returns quickly to its average value. Although normal fluctuations in energy facilitate the $\alpha$ response, the alignment inversion itself does not involve activation over an energy barrier. Additional evidence that $\alpha$ response is due to energy fluctuations, not activation, comes from analysis of the $T$ dependence of $\tau_\alpha$ using Eqs. (9) and (10), as in Sections 5.2 and 5.3, below.

Now return to the full simulation shown in Fig. 2 (A). Recall that the alignment ($m$, black line) exhibits two types of behavior, relatively rapid fluctuations around one orientation ($\beta$ response) combined with rare but abrupt inversions to the other orientation ($\alpha$ response). A similar bimodal distribution is deduced from NMR measurements [59,60]. Specifically, best agreement with the loss of angular correlation in glycerol is obtained using many (~98%) small-angle (~2°) fluctuations combined with rare (~2%) large-angle (~30°) jumps. From the open black circles in Fig. 2 (B), the average inversion process lasts ~500 MCS, yielding ~7,500 MCS for the 15 inversions in Fig. 2 (A), or ~0.6% of the total simulation. Although this fraction of time for inversions is influenced by $T$ and $N$, a bigger issue is that the OIM allows only 180° inversions. Thus, simulating ~30° jumps will require a more-detailed model. Nevertheless, a bimodal distribution arises in the OIM purely from equilibrium fluctuations of internal degrees of freedom, utilizing only simple and symmetrical constraints, with no bias in the local dynamics and no explicit long-time tails.

*5.2 Loss as a function of frequency*

Figure 3 is a log-log plot showing the out-of-phase (loss) component of response as a function of frequency. This loss is found by applying the fluctuation-dissipation theorem to power-spectral densities from equilibrium fluctuations in $m$, e.g. the black line in Fig. 2 (A). Figure 3 shows loss from simulations on systems containing (A) $N = 512$ and (B) $N = 27$ spins, at temperatures given in the figures. Several features characteristic of the dielectric loss in supercooled liquids can be identified in Fig. 3, including clear evidence for three types of response. The increase in $\chi''$ at highest frequencies involves microscopic dynamics from single spin flips. (Note that the microscopic frequency ($f_0$) is chosen so that the highest frequency gives $\log(f/f_0) \to 11$.) However, because microscopic dynamics is unrealistic in the Ising model, we focus on the other peaks that come from long-time thermal-equilibrium behavior.

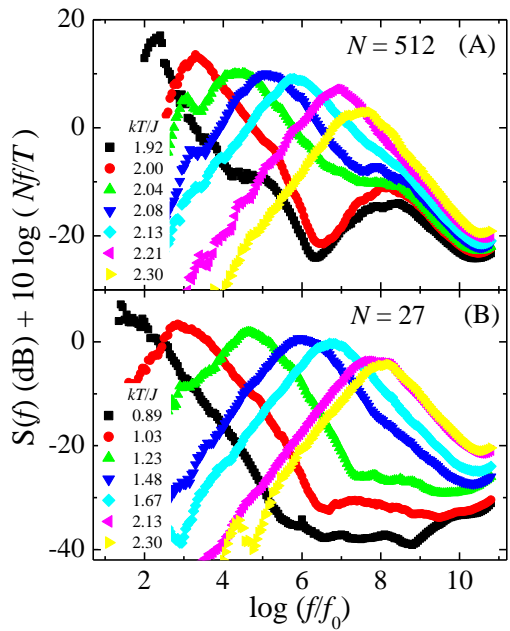

**Figure 3.** Frequency dependence of power spectral density (in dB) from time-dependent fluctuations in $m$, e.g. the black line in Fig. 2 (A). This $S(f)$ is converted to an out-of-phase (loss) component using the fluctuation-dissipation theorem by adding $10 \log(Nf/T)$, where 10 is needed for the logarithmic dB scale.

We identify the peak at lowest $f$ with the $\alpha$ response. It has the largest amplitude, and a super-Arrhenius shift towards lower $f$ as $T$ is reduced. It comes from alignment inversions, such as the sharp jumps in $m$ shown in Fig. 2 (A). Note that this $\alpha$ response is relatively narrow, having a FWHM of only about a decade, similar to single-exponential Debye-like relaxation. However, from Fig. 2 it is clear that this $\alpha$ response is not a smooth relaxation, instead involving sharp jumps with varying dwell times, so that Debye-like response arises only when averaged over all dwell times. A clear size dependence of $\tau_\alpha$ can be deduced by comparing Figs. 3 (A) and (B). Indeed, from theory, experiment, and simulations, response times are found to vary exponentially with inverse size. Therefore, when applied to an equilibrium distribution of region sizes [26], the $\alpha$ response becomes asymmetric, with an excess wing that extends to $f$ far above the $\alpha$ peak [23]. As for the amplitude of the $\alpha$ response, the fluctuation-dissipation theorem has an inherent $1/T$ dependence that dominates the amplitude of the loss peak, consistent with many measurements.

At low $T$, Fig. 3 (A) shows a peak at intermediate frequencies ($\log(f/f_0) \sim 8$) that we identify with the secondary ($\beta$) response. As in many measurements on real systems, this $\beta$ peak is broader than the $\alpha$ peak, with a simply-activated (Arrhenius-like) $T$ dependence. It comes from fluctuations in $m$ around either equilibrium alignment, as shown by the fast fluctuations between jumps in Fig. 2 (A). Thus, both $\alpha$ and $\beta$ responses come from the net alignment of all spins in the system, $m$, but their basic mechanisms are quite different: $\beta$ comes from normal fluctuations around relatively stable values, whereas $\alpha$ involves rare but abrupt inversions between these values. At the two lowest $T$ in Fig. 3 (A) there is a deep minimum between the $\alpha$ and $\beta$ peaks. This minimum indicates that the $\alpha$ response is suppressed as it approaches the frequency of the $\beta$ response, possibly from when local alignments cannot adapt fast enough to facilitate pathways through the entropy bottleneck.

Figure 3 (B) shows no clear $\beta$ peak from simulations on this small system, $N = 27$. Instead, there is a relatively flat valley at intermediate frequencies. This absence of a separate $\beta$ peak is consistent with many measurements, especially on substances with small internal systems (small IRR). For example, glycerol has $n \approx 18$ molecules at $T_g + 10$ K [11,12], with an excess high-$f$ wing on the $\alpha$ peak, but no separate $\beta$ peak.

*5.3 Temperature dependence of primary response times from measurements*

It is the super-Arrhenius $T$ dependence of the $\alpha$ response that gives the most stringent test of the OIM. For response spectra from simulations, such as those shown in Fig. 3, we define the $\alpha$-response time as the inverse of the $\alpha$-response frequency, $\tau_\alpha = 1/f_p$, where $f_p$ comes from fitting a Debye function, $\chi'' \propto f/\left[1 + (f/f_p)^2\right]$, to the primary response peak. (Again, we neglect a factor of $2\pi$ that simply shifts the behavior on a logarithmic scale.) An analogous procedure using the Havriliak-Negami function is applied to measured dielectric-loss spectra [52,53], yielding behavior that we will analyze in this section.

Open symbols in Fig. 4 show the $1/T$ dependence of $\tau_\alpha$ from measurements on five standard glass-forming liquids in (A) an Angell plot and (B) a modified Stickel plot. Solid lines in Fig. 4 come from fitting the data to Eq. (9) (the "VFT2 law"), which predicts straight lines (from Eq. (10)) when plotted as in Fig. 4 (B). In Fig. 4 (A), $T_g$ is defined by where $\tau_\alpha = 100$ s, and curvature indicates deviation from the Arrhenius law. For these measurements, only the $\beta$ response in sorbitol shows Arrhenius-like behavior (dot-dashed line). All other measurements show curvature characteristic of a super-Arrhenius $T$ dependence. Often, such curvature is attributed to the VFT law, but an exhaustive analysis shows clear deviations from the VFT law for most substances [55]. Another function, proposed by Mauro et al. (MYEGA [56]), is interesting because it has been shown to give better agreement than the VFT law for 7 out of 13 substances [57], and it has no divergence in $\tau_\alpha$ at finite $T$. Table I gives the $\chi^2$ values for these three functions. Quantitatively, from measurements on intermediate glass-forming liquids PG and glycerol, the VFT2 law gives $\chi^2$ values that are 1-2 orders of magnitude smaller than the other functions. Although fragile glass-forming liquids have comparable $\chi^2$ values for all three functions, linear behavior in Fig. 4 (B) is predicted by Eq. (10) only for the VFT2 law, qualitatively consistent with all measurements at $T_c/T < 0.6$ where the asymptotic mean-field approximation should be accurate.

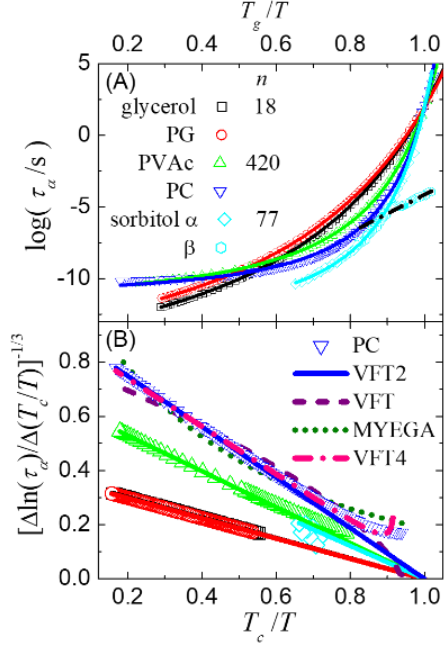

**Figure 4.** Inverse $T$ dependence of $\alpha$-response times in (A) an Angell plot and (B) a modified Stickel plot. Symbols show the behavior from measurements on five substances, listed in the legend, with data for sorbitol from Lunkenheimer et al. [57] and all other data from Stickel et al. [52-54]. (Abbreviations are listed in the caption for Table 1.) The legend also gives $n$, the number of molecules (or monomer units for PVAc) in a typical IRR at about 10 K above $T_g$, from available NMR measurements [11,12]. Beta response of sorbitol is shown by hexagons in (A), with the straight (dot-dashed) line from the Arrhenius law. Solid lines show fits to the $\alpha$ response using the VFT2 law, Eq. (9). Curvature in (A) is characteristic of a super-Arrhenius $T$ dependence. When plotted as in (B), the VFT2 law is linearized. On the scale of (B), the most conspicuous deviations from this VFT2 law occur in PC and PVAc as $T \rightarrow T_c$, where the high-$T$ expansion used for Eq. (9) is expected to fail. Including the quartic term from Eq. (8), the VFT4 function (dash-dotted line) shows improved agreement with the PC data, until $T_c/T > 0.8$ where higher-order corrections would be needed. Fits to these PC data using the VFT law (dashed) and MYEGA function (dotted) show continuous curvature across the entire range of $T_c/T$, unlike the data.

In Fig. 4 (B), the $T_c/T$-dependent differentials ($\Delta \ln(\tau_\alpha)/\Delta(T_c/T)$) from the data in Fig. 4 (A) are inverted, then raised to the 1/3 power, linearizing the VFT2 law. In contrast, the original Stickel plot had the same inverted differential, but raised to the 1/2 power, linearizing the VFT law. In Fig. 4 (B), all data (symbols) show linear behavior (solid lines) at $T_c/T < 0.6$, indicating clear qualitative agreement with the VFT2 law. Extrapolating these lines to zero (where the differential would diverge if mean-field theory remained valid) defines the mean-field critical temperature ($T_c$), similar to how the Weiss temperature in magnetism is defined by linear extrapolation of the Curie-Weiss law to zero (where the susceptibility would diverge). Even on the scale of Fig. 4 (B), propylene carbonate (PC, blue triangles) shows clear curvature, indicating systematic deviations from the VFT2 law. However, this curvature occurs at $T_c/T > 0.6$, where the quadratic term used for Eq. (9) is expected to fail. Indeed, the dot-dashed line in Fig. 4 (B) shows better agreement with data by adding the quartic term from Eq. (8) to Eq. (9) ("VFT4"), capturing the onset of deviations from the VFT2 law due to higher-order terms as $T_c/T \rightarrow 1$. Still, the PC data are clearly linear at $T_c/T < 0.6$, where the VFT2 law is expected to hold. Whereas, when plotted as in Fig. 4 (B) the VFT law shows curvature that is everywhere concave down, while the MYEGA formula shows curvature that is everywhere concave up, deviating qualitatively from all these data at $T_c/T < 0.6$.

| Liquid | $\chi^2$ (x100) | | | VFT2 fit parameters | | "molecules" | | Temperatures (K) | | | |
|---|---|---|---|---|---|---|---|---|---|---|---|
| | VFT2 | VFT | MYEGA | $C$ | $\log(\tau_\infty/s)$ | $n$ | $n/C$ | $T_c$ | $T_g$ | $\theta$ | $T_K$ |
| glycerol | 0.18 | 7.1 | 17 | 0.1069 | -17.745 | 18 | 170 | 104.1 | 191 | 129.6 | 135 |
| PG | 0.24 | 5.3 | 14 | 0.1035 | -17.242 | | | 90.61 | 170 | 113.5 | 127 |
| PVAc | 8.6 | 26 | 28 | 0.580 | -11.72 | 420 | 730 | 239.1 | 308 | 257.2 | 247 |
| PC | 50 | 56 | 19 | 1.636 | -11.24 | | | 150.4 | 164 | 144.5 | 127 |
| sorbitol | 33 | 31 | 41 | 0.45 | -13.3 | 77 | 170 | 193.6 | 267 | 233.4 | 226 |
| OTP | 6.9 | 6.8 | 7.2 | 0.072 | -30.6 | 35 | 480 | 151.8 | 283 | 175.6 | 200 |

**Table 1.** Parameters for six substances, from fitting the data shown in Fig. 4 [52-54,57] plus OTP data [61] (not shown). (Abbreviations: PG = propylene glycol, PVAc = polyvinyl acetate, PC = propylene carbonate, OTP = *o*-terphenyl.) Here, $n$ is the number of molecules (or monomer units for PVAc) in a typical IRR at $T_g + \sim 10\,K$, from correlation lengths measured by NMR [12] (with typical uncertainties of ≥30%) using the mass density and molecular mass. Kauzmann temperatures ($T_K$) are from [62] except for PVAc [63].

*5.4 Temperature dependence of primary response times from simulations*

Figure 5 shows temperature-dependent response times from simulations. As in Fig. 4, open symbols in Fig. 5 show the $1/T$ dependence of $\tau_\alpha$ in (A) an Angell plot and (B) a modified Stickel plot, from simulations on systems having six different sizes. Note several similarities between Figs. 5 and 4. Both have similar $\beta$ responses (hexagonal symbols in (A)) that can be characterized by the Arrhenius law (dot-dashed lines). Both have similar $\alpha$ responses, with curvature in (A) and intervals of linear behavior in (B) indicative of a super-Arrhenius $T$ dependence that obeys the VFT2 law (solid lines). Figure 5 (A) shows curvature (fragility) that increases with increasing $N$, while Fig. 4 (A) shows that fragility tends to increase with increasing $n$, at least for simple-molecule systems (the polymer, PVAc, is an outlier). Because this curvature is related to the slope when plotted as in (B), Fig. 5 (B) shows increasing magnitude of slope with increasing $N$. Similarly, Fig. 4 (B) shows a tendency of the magnitude of slope to increase with increasing $n$.

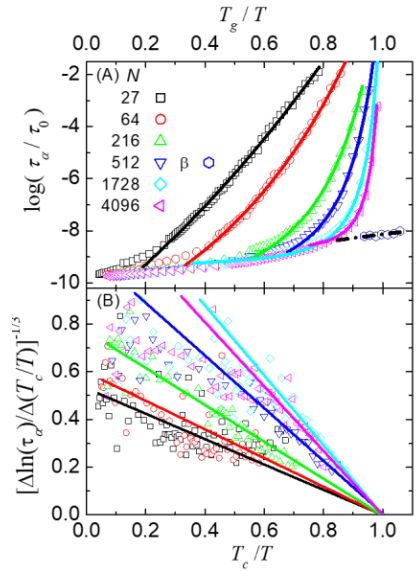

**Figure 5.** Inverse $T$ dependence of $\alpha$ response times in (A) an Angell plot and (B) a modified Stickel plot. Symbols show the behavior from simulations on systems of six sizes, listed in the legend. Beta response of the $N = 512$ system is given by the hexagons in (A), with a straight (dot-dashed) line from the Arrhenius law. Solid curves show best fits to the $\alpha$ response at high $T_g/T$ using the VFT2 law, Eq. (9). Curvature in (A) is characteristic of a super-Arrhenius $T$ dependence. When plotted as in (B) the VFT2 law is linearized, Eq. (10), with solid lines showing linear fits to the simulations at high $T_g/T$.

It is the distinct temperature regimes where the VFT2 law applies that reveal a crucial difference between measurements and simulations. Specifically, measurements in Fig. 4 (B) show best agreement with the VFT2 law at low $T_c/T$, with PC and PVAc exhibiting curvature as $T_c/T \to 1$. Whereas, Fig. 5 shows that simulations give best agreement with the VFT2 law at high $T_c/T$. Indeed, both Figs. 5 (A) and (B) show clear deviations from the solid lines at low $T_c/T$ for most values of $N$. We attribute these deviations to the failure of MC simulations to capture microscopic dynamics, and to simplifications in the OIM. For example, an isoenergetic spin flip in a simple-cubic lattice requires that the given spin has exactly 0, 2, 4, or 6 interacting neighbors with half of these neighbors up. In contrast, Figs. 4 (A) shows that the VFT2 law gives good agreement with measurements of $\tau_\alpha$, even at highest $T$. We attribute this success to the myriad of local environments in real amorphous systems, combined with other mechanisms for conservation of energy (such as vibrational energies) that are not included in the OIM.

*5.5 Size of independently relaxing regions from primary response*

Simulations of systems as a function of size allow characterization of size-dependent behavior, which can be extended to experiments. From the $T_c/T$ scaling used in the modified Stickel plot, Eq. (10) predicts that both the magnitude of the slope (|slope|) and the intercept as $T_c/T \to 0$ should be $(C/2)^{1/3} = (2N)^{1/3}$. Figure 6 shows results from simulations on systems having ten different values of $N$, and from our analysis of measurements on the four liquids where $n$ has been measured directly by NMR [11,12]. For the three simple-molecule liquids, a fit using $|\text{slope}| = An^B$ (solid red line) yields $A = 0.14\pm0.02$ and $B = 0.32\pm0.10$. Similarly, a fit to simulations with $N < 1730$ (solid black line) yields $|\text{slope}| = AN^B$, with $A = 0.22\pm0.06$ and $B = 0.26\pm0.05$. Given the relatively large uncertainties, experiments and simulations are consistent with the theoretically expected exponent, $B = 1/3$. However, the amplitudes ($A$) do not agree with the expected $2^{1/3} = 1.2599$. At least part of this discrepancy comes from the assumption that $C$ in Eq. (9) is independent of $T$. Hidden $T$ dependences in Eq. (9) include: $T_c$ from the changing fraction of interacting bonds, $n$ from measured changes in IRR sizes, and fluctuations that do not reach $E = 0$ for the $\alpha$ response. For example, the red lines in Fig. 2 (B) show that on average, $\varepsilon/J$ fluctuates only about 2/3 of the way to zero. Nevertheless, from the behavior shown in Fig. 6 we argue that $\alpha$-response measurements as a function of $T$ can be used to estimate the size of IRR in simple-molecule glass-forming liquids.

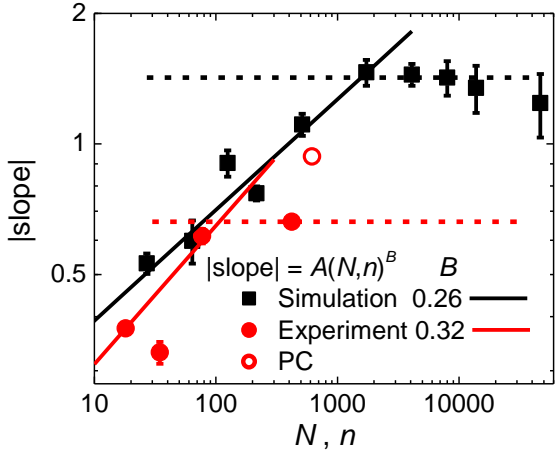

**Figure 6.** Log-log plot of slopes found from modified Stickel plots. The magnitude of these slopes is plotted as a function of the system size ($N$) for simulations (black), or size of IRR ($n$) for experiments (red) [12]. Simulations (squares) are from Fig. 5 (B) and experiments (circles) from Fig. 4 (B), plus OTP (circle with largest error bar) that is not shown in Fig. 4 because its weak dielectric response yields a large uncertainty. Equations (9) and (10) predict a slope of $B=1/3$. The open red circle has $n = 620$, estimated from VFT4 fits to the PC data in Fig. 4 (B), as there are no NMR measurements of $n$ for this substance.

From the behavior shown in Fig. 6, |slope| increases with increasing simulation size until $N\sim 1730$, where |slope| saturates to a value of 1.4 (dashed line). Although experimental results on simple-molecule systems show a general trend of increasing |slope| with increasing $n$, the polymer (PVAc) does not follow this trend, having |slope|=0.662±0.001 (dashed line). Thus, for PVAc, predicting the size of IRR from dielectric measurements is more difficult. One possibility is that |slope| saturates to the value of PVAc, similar to the saturation seen for the simulations. However, this seems unlikely given the large value of |slope| for PC in Fig. 4 (B). Indeed, from the VFT4 fit to the PC data we deduce that PC has $n = 620\pm50$ molecules, as shown by the open circle in Fig. 6. Therefore, we speculate that response in polymers has a different dependence on IRR size than in simple glass-forming liquids. For example, if the fluctuations used to derive Eq. (9) come from monomer units, not separate molecules, a different dependence on size might be expected. Furthermore, because OTP (solid circle with largest error bar in Fig. 6) also falls outside the overall trend, additional studies will be necessary to confirm the $C \propto n$ dependence of experiments.

*5.6 Hysteresis as a function of temperature*

Figure 7 shows three ways of representing the cooling- and heating-rate dependence of the OIM as a function of $kT/J$: (A) gives $\bar{\varepsilon}/J$, (B) its difference between cooling and heating $(\bar{\varepsilon}_- - \bar{\varepsilon}_+)/J$, and (C) the specific heat $c_{V\pm} = \Delta\bar{\varepsilon}/(k\Delta T_\pm)$. All results come from the energy per particle averaged over the entire simulation run at each temperature, $\bar{\varepsilon} = \bar{E}/N$. Here, $\bar{E}$ is the enthalpy of the OIM because magnetic field and pressure are zero, with volume (*V*) fixed. The simulations start at an initial temperature of $kT_0/J = 2.40$ (off scale to the right). Steps down in *T* use a constant factor (*a*), yielding a variable step size $\Delta T_- = aT_r - T_r$ (the subscript on $\Delta T$ denotes its sign). The minimum temperature, as determined by having 20 steps for $a = 0.92$, is $kT_{min}/J = 0.92^{20}kT_0/J = 0.453$ (off scale to the left). Steps back up to $T_0$ use the inverse factor $1/a$, yielding variable step size $\Delta T_+ = T_r/a - T_r$. The other two constant factors, as given in the legend of Fig. 7 (C), are $a = 0.92^2 = 0.8464$ (15.4 %) and $0.92^{1/2} = 0.9592$ (4.1 %); thus, all step sizes share some common temperatures.

Simulations in Fig. 7 utilized $P = 1$, so that $\bar{\varepsilon}$ is averaged over $Q = \tau \times 10^P \approx 1.31M$ MCS. Additional averaging, especially important for the differences in Figs. 7 (B) and (C), is achieved without changing $Q$ by repeating each cooling and heating cycle at least 16 times. We identify a hysteresis temperature, $kT_h/J \approx 1.4$ and 1.8 for $N = 27$ and 1728, respectively. Below this $T_h$, the averaging time $Q$ is too short to fully explore all aspects of the behavior, primarily due to the rarity of spin inversions ($\alpha$ response). Indeed, because $Q$ corresponds to $\log(f/f_0)\sim 5$, the freezing of $\alpha$ response below $kT_h/J \approx 1.4$ for $N = 27$ is consistent with Fig. 3. However, Fig. 7 (A) shows that $\bar{\varepsilon}$ continues to decrease with decreasing *T* until significantly below $T_h$, a consequence of the increasing density of low-energy bonds that are favored at low *T*. The OIM has no contribution from vibrational energy, hence there is no underlying Debye-like ($T^3$) specific heat, but other features in Fig. 7 mimic the hysteresis measured around $T_g$ in most glass-forming liquids [64]. For example, $c_{V-}$ has a gradual step down upon cooling, while $c_{V+}$ has a more-rapid step up, with steepness and overshoot that increase with decreasing rate of temperature change. Also, $T_h$ tends to shift to lower *T* with decreasing $|\Delta T_\pm|$, and $(\bar{\varepsilon}_- - \bar{\varepsilon}_+)/J$ increases with decreasing *N*. Although experimental values of $T_g$ are often near the midpoint of the hysteresis [64], using $T_h$ for the onset of hysteresis can be useful to identify were the $\alpha$ response freezes on a given time scale.

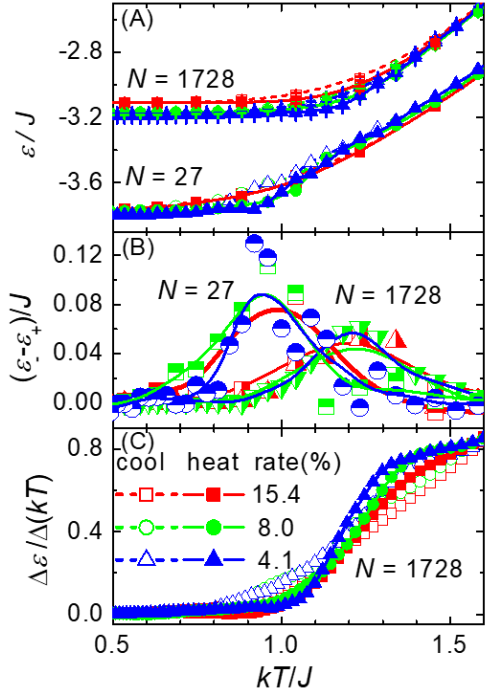

**Figure 7.** (A) Normalized average energy per spin, (B) its difference between cooling and heating, and (C) specific heat. Data are from simulations of systems with two different sizes (except in (C)) and three different temperature change rates, as given in the legend of (C). The data in (C) are smoothed, with cubic-spline interpolation for clarity. The behavior in (B) can be used to estimate the temperatures below which hysteresis appears: $kT_h/J \approx 1.4$ and $1.8$ for $N = 27$ and $1728$, respectively.

## 6. Discussion

### 6.1 Summary of results from the OIM

The OIM is based on three assumptions not found in most previous models of the liquid-glass transition. These are: explicit finite-size effects from independent small systems, neighboring particles that might not interact, and orthogonal dynamics that allows energy and alignment to fluctuate independently. Direct evidence showing different time scales for dipole rotation and energy flow comes from nonlinear dielectric measurements at frequencies far above the dielectric loss peak, $f \gg f_p$, where dozens of pump oscillations are required before energy is equilibrated [29-32]. Justification for independent small systems comes from several experimental techniques showing that dynamic heterogeneity dominates the $\alpha$ response of glass-forming liquids [6-14]. Justification for allowing neighboring spins that might not interact comes from the increase in entropy that yields an equilibrium distribution of interaction energies [26], and/or from a distribution of local environments that may intermittently interrupt the interactions between particles. We anticipate that in more-sophisticated models of interacting molecules there will be a higher cost in energy to form isolated non-interacting bonds, which will favor forming continuous interfaces surrounding relatively compact regions. We speculate that these interfaces will define a break in the quantum coherence between distinct wavefunctions, yielding independently relaxing regions. Thus, it is likely that two of the assumptions in the OIM are connected, and that more-detailed models will develop their own equilibrium distribution of IRR. Indeed, using the nanocanonical ensemble, theoretical expressions for the equilibrium distributions of small systems have been found for the 1-D Ising model with intermittent interactions and for a semi-classical ideal gas that yields a novel solution to Gibbs' paradox [26].

In Section 5 we have shown that the OIM mimics more than twenty characteristics commonly found in glass-forming liquids [1,2]. Eight of these characteristics are found in the *T*-dependence of the average energy, Fig. 7. Specifically, as *T* is reduced through $T_h$ we find: 1) a change in slope of $\bar{\varepsilon}$ that yields 2) a gradual step down in $c_{V-}$. As *T* is increased through $T_h$ we find 3) hysteresis that yields 4) a sharper step up in $c_{V+}$. Furthermore, if the rate of heating through $T_h$ is decreased we find: that 5) $T_h$ is decreased and 6) the step up in $c_{V+}$ is sharpened, yielding 7) increased overshoot ($c_{V+} > c_{V-}$). In addition, Fig. 7 (B) shows that the magnitude of the hysteresis is smaller for larger *N* (and hence for higher fragility), consistent with 8) measurements summarized in Ref. [64]. Figure 5 (A) shows that the 1/*T* dependence of the OIM mimics three common characteristics of supercooled liquids. We find *β* response that exhibits: 9) Arrhenius activation, in addition to *α* response with 10) curvature consistent with super-Arrhenius *T* dependence and 11) increasing curvature with increasing *N*, covering a range of fragilities similar to the data shown in Fig. 4 (A). Figure 3 shows that as a function of frequency at fixed *T*, the OIM mimics seven features found in the response of glass-forming liquids. We find: 12) a partial peak at highest *f* due to microscopic processes, 13) a *β*-response peak at intermediate *f* that is 14) relatively broad, 15) symmetrical, and 16) suppressed in systems with small *N*. In addition, for the *α*-like primary response peak at lowest *f* we find: 17) an amplitude that increases roughly as 1/*T*, 18) a width consistent with single-exponential relaxation for systems of a single size, and 19) relaxation times that vary exponentially with inverse *N*, which will yield asymmetrical primary response peaks in systems having a thermal equilibrium distribution of *N*. Figure 2 shows that as a function of *t*, the OIM exhibits two additional characteristics of glass-forming liquids. From Fig. 2 (A) we find that 20) the primary response fluctuates around one alignment, then abruptly jumps to fluctuate around the other alignment, reminiscent of the behavior deduced from NMR measurements; and from Fig. 2 (B) we find that 21) there is a coincidence between fluctuations that increase energy and the primary response from alignment inversion. Although this coincidence seems to suggest that an energy increase is needed for activation over a barrier, careful analysis indicates that primary response in the OIM comes from increased entropy needed for pathways through a bottleneck. Thus, despite its simplicity, the OIM mimics many properties of supercooled liquids, and provides new understanding of the liquid-glass transition.

Further insight comes from a theoretical analysis of the OIM. Indeed, in section 4, mesoscopic mean-field theory is utilized to predict some novel aspects of glass-forming liquids that are consistent with results found in section 5. In fact, a key component of the OIM is that finite-size effects enhance the energy fluctuations, especially in mean-field theory. These energy fluctuations yield an expression for the *T* dependence of *α*-response times, the "VFT2 law." This VFT2 law diverges inversely proportional to the square of the difference in *T* from a critical temperature, in contrast to the linear divergence of the VFT law. The VFT2 law is linearized using a modified Stickel, Fig. 4 (B). All data presented in Fig. 4 (B) show agreement with the VFT2 law at high *T*, where mean-field theory is expected to hold. When plotted as in Fig. 4 (B), other proposed expressions such as the VFT law and MYEGA formula show curvatures that are qualitatively inconsistent with the data when the entire range of *T* is considered. Furthermore, mean-field theory on the OIM predicts response times that vary exponentially with inverse size. When applied to the expected distribution of IRR inside bulk samples, this size dependence is known to yield an asymmetric peak [47], and improved agreement with the deduced distributions of relaxation times [50,51]. Furthermore, this size dependence can be used to deduce the size of the IRR, as shown in Fig. 6.

*6.2 Comparison to some other models of glass-forming liquids*

A stringent test of any model for supercooled liquids is to mimic measured super-Arrhenius *T* dependences in the *α* response. Many models are based on mechanisms for divergent activation energies that yield the VFT law. In contrast, the OIM predicts a novel *T* dependence for $\tau_\alpha$, the "VFT2 law," arising from energy fluctuations that facilitate pathways through an entropy bottleneck, not activation over an energy barrier. Nevertheless, other aspects of the OIM are similar to previous models, which we now discuss.

Two of the earliest models applied to glass-forming liquids are the free-volume [65,66] and defect-diffusion [67,68] pictures. The intermittent bonds in the OIM simulate some aspects of these pictures. Specifically, non-interacting bonds are a type of defect that diffuse through the lattice, facilitating primary response (spin flips) as they diffuse, similar to the mechanism of free volume. Indeed, spin flips occur only if half of the interacting neighbors are up and the other half are down, a spin-alignment version of a soft molecular environment [69,70]. However, in the OIM the freezing of non-interacting bonds yields the hysteresis below $T_h$, as shown in Fig. 7, which is peripheral to the primary response that yields the VFT2 law and the diverging time scales as $T \to T_c$. Furthermore, none of these earlier pictures gives an underlying phase transition. Still, an important future step will be to extend ideas from the OIM to more-detailed models incorporating molecular motion and elasticity.

Another early model that yields divergent time scales around $T_g$ is the Adam-Gibbs picture of an activation energy that varies inversely proportional to configurational entropy [71]. Configurational entropy also plays an important role in the OIM. However, in the OIM primary response is attributed to energy fluctuations that increase the pathways through an entropic bottleneck, not activation over an energy barrier. Furthermore, NMR measurements yield IRR containing $n$~10-100 molecules, quantitatively consistent with the behavior of the OIM as shown in Fig. 6, unlike the 4-8 molecules deduced from the Adam-Gibbs picture [72,73]. Nevertheless, the success of the OIM supports the notion that configurational entropy is central to supercooled liquids.

Both mode-coupling theory (MCT) [74,75] and the OIM attribute the behavior of supercooled liquids to an underlying transition. Furthermore, MCT is usually treated using mean-field theory, which also provides a useful approximation to the OIM (see Section 4.1). However, MCT involves an "avoided" dynamical transition, whereas $T_c$ in the OIM is from a thermal transition that is smeared out by finite-size effects. Furthermore, the critical temperature in MCT (where key dynamical changes occur) is above $T_g$, whereas $T_c < T_g$ in the OIM. Thus, high-$T$ mean-field expressions, such as the VFT2 law, can remain relatively accurate down to $T_g$, especially in relatively strong glass-forming liquids as shown by the solid lines in Fig. 4. Another connection to MCT may come from simulations of the OIM, Fig. 5 (A), where there are clear deviations from the VFT2 law at high-$T$. We attribute these deviations to the specific local configurations needed for isoenergetic spin flips in the simple-cubic lattice of the OIM. Although no such deviations are seen in the dielectric data of Fig. 4, other measurement techniques with stronger coupling to local dynamics show evidence for a crossover that may be related to the behavior seen in simulations of the OIM.

The OIM also shares some similarities with a random first-order transition (RFOT) [76]. Both attribute the behavior of supercooled liquids to an underlying thermal transition at $T < T_g$. The transition of the RFOT occurs at the Kauzmann temperature $T_K$, where configurational entropy is extrapolated to reach zero. Although the novel $T$-dependence of the VFT2 law yields a somewhat different extrapolation, $T_c$ in the OIM is often near to $T_K$ (see Table I). However, $T_c$ comes from a high-$T$ mean-field extrapolation, so that finite-size fluctuations will suppress the thermal transition to below $T_c$, similar to non-classical critical scaling in ferromagnets [23,77]. As its name implies, the RFOT has a discontinuous jump in the order parameter at $T_K$, but there is also a gradual increase in order around $T_K$ so that the RFOT is second order in the Ehrenfest sense. The OIM has no discontinuous jump, exhibiting only a gradual onset of order from the Ising model that starts as a second-order transition and is broadened by finite-size effects. In fact, these finite-size effects yield an order parameter (average magnitude of alignment) that is nonzero at all $T$, with only an inflection point near the center of the thermal transition. In general, the RFOT is treated using mean-field theory, which is also a useful way to approximate the OIM, yielding e.g. the VFT2 law. Some studies suggest that the RFOT transition becomes unstable outside of mean-field theory [78,79], whereas simulations of the OIM utilize microscopic interactions to mimic many features of liquid-glass behavior. Moreover, the OIM is approximated using mesoscopic mean-field theory, with finite-size effects that are essential for energy fluctuations that open pathways through the entropy bottleneck yielding the $\alpha$ response.

Both MCT and the RFOT theory attribute slow dynamics in supercooled liquids to activation over energy barriers in a high-dimensional energy landscape. The energy landscape is assumed to form far above $T_g$, with the system becoming increasingly trapped in energy minima of the landscape as $T \to T_g$. Thus, these theories have complexity in the multi-dimensional configurational space of the landscape. In these theories, and others [80,81], heterogeneity is assumed to come from quenched disorder that is frozen into the system. In contrast, heterogeneity in the OIM is assumed to come from the nanocanonical ensemble in nanothermodynamics, which yields a heterogeneous distribution of IRR in thermal equilibrium. Thus, each local region of the OIM has a single potential-energy barrier in its one-dimensional alignment space, with the complexity arising in real space, consistent with measurements of dynamic heterogeneity [6-14]. Furthermore, in the OIM, slow relaxation involves finding pathways through an entropy bottleneck, not activation over an energy barrier. In a more-detailed model, non-interacting bonds should form continuous interfaces between the regions, facilitating the thermal equilibrium distribution of IRR in the nanocanonical ensemble.

Frustration-based models attribute the slow relaxation to an avoided thermodynamic critical point far above $T_g$ [82]. Below this critical point, frustration from an inability to match local and global structures yields a nonequilibrium mosaic of configurations that are frozen into the sample. A specific example is the frustration-limited domain (FLD) model [80,81]. The OIM is also based on heterogeneous regions inside the sample. However, in the OIM these IRR are identified by their dynamics (not structure), with neighboring regions having uncorrelated fluctuations, consistent with several experimental techniques [6-14], and as needed for their entropies to be additive. Furthermore, these regions are assumed to be in thermal equilibrium whenever $T > T_g$, with a distribution of sizes from the nanocanonical ensemble, consistent with measurements indicating a change in the distribution at $T_g$ [50]. In the FLD picture, the fragility of supercooled liquids (curvature on an Arrhenius plot) varies inversely with the degree of frustration, so that the curvature increases monotonically with the length scale of cooperativity. Similarly, in the OIM there is an increase in curvature with increasing region size, *n* or *N*, as shown in Figs. 4 (A) and 5 (A). Indeed, this curvature yields the magnitude of slope that is found to follow the cube-root size dependence expected from mean-field theory, as shown in Fig. 6.

The OIM combines components from mesoscopic mean-field theory [22] and an Ising model with entropic constraints [83]. Specifically, in section 4.2, the OIM is approximated using finite-size mean-field theory, but with energy fluctuations that facilitate pathways through an entropy bottleneck yielding the VFT2 law. The Ising model with entropic constraints that was previously used to simulate supercooled liquids and ferromagnets [23] uses a local entropy bath to maintain maximum entropy, with a bypass mechanism for spins that have no net energy change. Thus, this bypass mechanism is the isoenergetic step in the orthogonal dynamics of the OIM. These previous models yield the VFT law and stretched-exponential relaxation, but lack the full range of liquid-glass behavior found from the OIM.

A 1-D version of the OIM was previously used to mimic the 1/*f*-like noise from a qubit [26]. We implement three main changes to that model. First, we extend the OIM to 3-D, yielding a phase transition and matching the dimensionality of most samples. Second, we remove the local entropy bath, as expected for distinguishable particles and nondegenerate states due to the variety of local environments in amorphous systems. Third, the OIM described here has intermittent interactions between spins, driven by the resulting increase in entropy and variable local environments, which allows the system to attain an equilibrium distribution of interaction energies.

## 7. Conclusions

Here we study the thermal and dynamic properties of an Ising model with novel constraints. This orthogonal Ising model (OIM) treats finite-size systems using orthogonal dynamics, with intermittent interactions between spins. Orthogonal dynamics separates conservation of energy from conservation of alignment, allowing these fundamental laws to evolve independently on their own preferred time scales; while the intermittent bonds yield a thermal distribution of interaction energies. The OIM mimics more than twenty characteristics that are commonly found in supercooled liquids and glasses, as summarized in Section 6.1. From the OIM we deduce that the liquid-glass behavior is due to an underlying $2^{nd}$-order phase transition that is broadened by finite-size effects. Perhaps the most stringent test of the OIM comes from the peak response time of supercooled liquids as a function of $1/T$, shown in Figs. 4-5. In section 4.1, a mean-field approximation to the OIM yields $\tau_\alpha \propto \exp\{1/[C(1 - T_c/T)^2]\}$. Here, the critical temperature ($T_c$) is where the mean-field transition would occur if extrapolated from high $T$, while $C$ is related to the curvature on an Arrhenius-like plot (fragility) and is generally proportional to system size. The mean-field expression for $\tau_\alpha$ is reminiscent of the VTF law, but with the temperature difference in the denominator squared, so that we call it the VFT2 law. Figure 4 (B) shows measurements of several glass-forming liquids plotted in a modified Stickel plot that linearizes the VFT2 law. This plot shows linear behavior for all substances at high $T$, where mean-field theory is expected to hold. Such qualitative consistency with the VFT2 law cannot be matched by the VFT law, or other functions previously used for $\tau_\alpha$ in glass-forming liquids.

As a function of time, Fig. 2 shows that the alignment of the OIM exhibits two types of response: fast fluctuations around one alignment, with relatively rare but sudden inversions to the other alignment. We associate these fluctuations with the $\beta$ and $\alpha$ responses, respectively. The $\beta$ response shows a relatively broad peak (Fig. 3 (A)) with Arrhenius-like activation (Figs. 4 (A) and 5 (A)), while the $\alpha$ response exhibits super-Arrhenius behavior. A key result from simulations and mean-field theory of the OIM is that this $\alpha$ response comes from energy fluctuations that enhance the possible pathways through an entropy bottleneck, not activation over an energy barrier. The dependence of the $\alpha$ response on system size is consistent with the distribution of relaxation times deduced from measured relaxation in many systems. The sizes of IRR found from response measurements using the OIM agrees with the sizes measured directly by NMR.

By adapting the simplest microscopic picture for a thermodynamic phase transition, we find that a finite-size Ising model with orthogonal dynamics and intermittent interactions mimics more than twenty distinctive features found in supercooled liquids and the glass transition. Despite its simplicity, this orthogonal Ising model provides a novel framework for interpreting the behavior of glass-forming liquids, and a foundation for developing more-detailed models.

**Funding:** This research received no external funding

**Acknowledgments:** I acknowledge support from Arizona State University Research Computing for use of their facilities. I thank Sumiyoshi Abe, Roland Böhmer, Peter Lunkenheimer, Vladimiro Mujica, and Ranko Richert for their careful reading of the manuscript, and many constructive comments. I also thank Ranko Richert and Peter Lunkenheimer for providing the original data presented here.

**Conflicts of Interest:** The author declares no conflict of interest.

**References**

1. Angell, C.A. Ten questions on glassformers, and a real space 'excitations' model with some answers on fragility and phase transitions, *J. Phys. Condens. Matter* **2000**, *12*, 6463-6475.

2. Dyre, J.C. Ten themes of viscous liquid dynamics, *J. Phys. Condens. Matter* **2007**, *19*, 205105.

3. Niss, K.; Hecksher, T. Perspective: Searching for simplicity rather than universality in glass-forming liquids, *J. Chem. Phys.* **2018**, *149*, 230901.

4. Brush, S.G. History of the Lenz-Ising model, *Rev. Mod. Phys.* **1967**, *39*, 883-893.

5. Niss, M. History of the Lenz-Ising model 1920-1950: from ferromagnetic to cooperative phenomena, *Arch. Hist. Exact Sci.* **2005**, *59*, 267-318.

6. Donth, E. The size of cooperatively rearranging regions at the glass transition, *J. Non-Cryst. Solids* **1982**, *53*, 325-330.

7. Yukalov, V.I. Phase transitions and heterophase fluctuations, *Phys. Rep.* **1991**, 208, 395-489.

8. Böhmer, R.; Chamberlin, R.V.; Diezemann, G.; Geil, B.; Heuer, A.; Hinze, G.; Kuebler, S.C.; Richert, R.; Schiener, B.; Sillescu, H.; Spiess, H.W.; Tracht, U.; Wilhelm, M. Nature of the non-exponential primary relaxation in structural glass-formers probed by dynamically selective experiments, *J. Non-Cryst. Solids* **1998**, *235-237*, 1-9.

9. Ediger, M.D. Spatially heterogeneous dynamics in supercooled liquids, *Annu. Rev. Phys. Chem.* **2000**, *51*, 99-128.

10. Richert, R. Heterogeneous dynamics in liquids: fluctuations in space and time, *J. Phys. Condens. Matter* **2002**, *14*, R703-R738.

11. Reinsberg, S.A.; Heuer, A.; Doliwa, B.; Zimmermann, H.; Spiess, H.W. Comparative study of the NMR length scale of dynamic heterogeneities of three different glass formers, *J. Non-Cryst. Sol.* **2002**, *307-310*, 208-214.

12. Qiu, X.; Ediger, M.D. Length scale of dynamic heterogeneity in supercooled D-sorbitol: Comparison to model predictions, *J. Phys. Chem. B* **2003**, *107*, 459-464.

13. Berthier, L.; Biroli, G.; Bouchaud, J.P.; Cipelletti, L.; El Masri, D.; L'Hôte,; Ladieu, F.; Pierno, M. Direct experimental evidence of a growing length scale accompanying the glass transition, *Science* **2005**, *310*, 1797-1800.

14. Kaufman, L.J. Heterogeneity in single-molecule observables in the study of supercooled liquids, *Ann. Rev. Phys. Chem.* **2013**, *64*, 177-200.

15. Kac, M. (1971): "The role of models in understanding phase transitions," in Mills, Ascher, and Jaffee (1971), pp. 23–39.

16. Niss, M. History of the Lenz-Ising model 1950-1965: from irrelevance to relevance, *Arch. Hist. Exact Sci.* **2009**, *63*, 243-287.

17. Kawasaki, K. Diffusion constants near the critical point for time-dependent Ising models. I, *Phys. Rev.* **1966**, *145*, 224-230.

18. Fredrickson, G.H.; Andersen, H.C. Kinetic Ising model of the glass transition, *Phys. Rev. Lett.* **1984**, *53*, 1244-1247.

19. Chamberlin, R.V.; Stangel, K.J. Monte Carlo simulation of supercooled liquids using a self-consistent local temperature, *Phys. Lett. A* **2006**, *350*, 400-404.

20. Fredrickson, G.H.; Brawer, S.A. Monte Carlo investigation of a kinetic Ising model of the glass transition, *J. Chem. Phys.* **1986**, *84*, 3351-3366.

21. Berthier, L.; Biroli, G. Theoretical perspective on the glass transition and amorphous materials, *Rev. Mod. Phys.* **2011**, *83*, 587-645.

22. Chamberlin, R.V. Mesoscopic mean-field theory for supercooled liquids and the glass transition, *Phys. Rev. Lett.* **1999**, *82*, 2520-2523.

23. Chamberlin, R.V. The big world of nanothermodynamics, *Entropy* **2015**, *17*, 52-73.

24. Fichthorn K.A.; Weinberg, W.H. Theoretical foundations of dynamical Monte Carlo simulations, *J. Chem. Phys.* **1991**, *95*, 1090-1096.

25. Voter, A.F.; Montalenti, F.; Germann, T.C. Extending the time scale in atomistic simulations of materials, *Annu. Rev. Mater. Res.* **2002**, *32*, 321-326.

26. Chamberlin, R.V.; Clark, M.R.; Mujica, V.; Wolf, G.H. Multiscale thermodynamics: Energy, entropy, and symmetry from atoms to bulk behavior *Symmetry* **2021**, *13*, 721.

27. Häggkvist, P.; Rosengren, A.; Lundow, P.H.; Markström, K.; Andrén, D.; Kundrotas, P. On the Ising model for the simple cubic lattice, *Adv. Phys.* **2007**, *56*, 653-755.

28. Newman, M.E.J.; Barkema, G.T. *Monte Carlo Methods in Statistical Physics*, Clarendon Press, Oxford, UK, **1999**.

29. Schiener, B.; Böhmer, R.; Loidl, A.; Chamberlin, R.V. Nonresonant spectral hole burning in the slow dielectric response of supercooled liquids, *Science* **1996**, *274*, 752-754.

30. Schiener, B.; Chamberlin, R.V.; Diezemann, G.; Böhmer, R. Nonresonant dielectric hole burning spectroscopy of supercooled liquids; *J. Chem. Phys.* **1997**, *107*, 7746-7761.

31. Weinstein, S.; Richert, R. Heterogeneous thermal excitation and relaxation in supercooled liquids, *J. Chem. Phys.* **2005,** *124*, 224506.

32. Chamberlin, R.V.; Böhmer, R.; Richert, R. Nonresonant spectral hole burning in liquids and solids. In *Nonlinear dielectric spectroscopy*, Richert, R., Ed.; Springer, Cham, Switzerland, **2018**, pp. 127-185.

33. Chang, I.; Fujara, F.; Geil, B.; Heuberger, G.; Mangel, T.; Sillescu, H. Translation and rotational molecular motion in supercooled liquids studied by NMR and forced Rayleigh scattering; J. Non-Cryst. Solds. **1994**, *172-174*, 248-255.

34. Edmond K.V.; Elsasser, M.T.; Hunter, G.L.; Pine, D.J.; Weeks, E.R. Decoupling of rotational and translational diffusion in supercooled colloidal fluids, *Proc. Nat. Acad. Sci.* **2012**, 109, 17891-17896.

35. Chamberlin, R.V.; Mujica, V.; Izvekov, S.; Larentzos, J.P. Energy localization and excess fluctuations from long-range interactions in equilibrium molecular dynamics, *Physica A* **2020**, *540*, 123228.

36. Hänggi, P.; Talkner, P.; Borkovec, M. Reaction-rate theory: fifty years after Kramers, *Rev. Mod. Phys.* **1990**, *62*, 251-341.

37. Zhou, H.X.; Zwanzig, R. A rate process with an entropy barrier, *J. Chem. Phys.* **1991**, *94*, 6147-6152.

38. Pedone, D.; Langecker, M.; Abstreiter, G.; Rant, U. A pore-cavity-pore device to trap and investigate single nanoparticles and DNA molecules in a femtoliter compartment: confined diffusion and narrow escape, *Nano Lett.* **2011**, *11*, 1561-1567.

39. Cofer-Shabica, D.V.; Stratt, R.M. What is special about how roaming chemical reactions traverse their potential surfaces? Differences in geodesic paths between roaming and non-roaming events, *J. Chem. Phys.* **2017**, *146*, 214303.

40. Landau, L.D.; Lifshitz, E.M. *Statistical Physics 2nd ed.*; Pergamon Press, Oxford, UK, **1969**.

41. Pathria R.K.; Beale, P. *Statistical Mechanics 3rd ed.*; Elsevier, Oxford, UK, **1996**.

42. Gradshtein I.S.; Ryzhik, I.M. *Tables of integrals, series and products*, ed. by Jeffrey, A.; Zwillinger, D. Academic Press, Burlington, MA, USA, **2007**.

43. *Handbook of mathematical functions with formulas, graphs, and mathematical tables*, ed. by Abramowitz M.; Stegun, I.A. National Bureau of Standards, Applied Mathematics Series, 55, US Government Printing Office, Washington DC, USA, **1967**.

44. Hall, R.W.; Wolynes, P.G. The aperiodic crystal picture and free energy barriers in glasses, *J. Chem. Phys.* **1987**, *86*, 2943-2948.

45. Drozd-Rzoska, A.; Rzoska, S.J.; Starzonek, S. New paradigm for configurational entropy in glass-forming systems, *Sci. Rep.* **2022**, *12*, 3058.

46. Bässler. H. Viscous flow in supercooled liquids analyzed in terms of transport theory for random media with energetic disorder, *Phys. Rev. Lett.* **1987**, *58*, 767-770.

47. Chamberlin, R.V.; Haines, D.N. Percolation model for relaxation in random systems, *Phys. Rev. Lett.* **1990**, *65*, 2197-2200.

48. Chamberlin, R.V.; Holtzberg, F. Remanent magnetization of a simple ferromagnet, *Phys. Rev. Lett.* **1991**, *67*, 1606-1609.

49. Chamberlin, R.V.; Scheinfein, M.R. Slow magnetic relaxation in iron: A ferromagnetic liquid, *Science* **1993**, *260*, 1098-1101.

50. Chamberlin, R.V.; Böhmer, R.; Sanchez, E.; Angell, C.A. Signature of ergodicity in the dynamic response of amorphous systems, *Phys. Rev. B* **1992**, *46*, 5787-5790.

51. Hansen, C.; Richert, R.; Fischer, E.W. Dielectric loss spectra of organic glass formers and Chamberlin cluster model, *J. Non-Cryst. Solids* **1997**, *215*, 293-300.

52. Stickel, F.; Fischer, E.W.; Richert, R. Dynamics of glass-forming liquids. I. Temperature derivative of dielectric relaxation data, *J. Chem. Phys.* **1995**, *102*, 6251-6257.

53. Stickel, F.; Fischer, E.W.; Richert, R. Dynamics of glass-forming liquids. II. Detailed comparison of dielectric relaxation, dc-conductivity, and viscosity data, *J. Chem. Phys.* **1996**, *104*, 2043-2055.

54. Stickel, F. Untersuchung der Dynamik in niedermolekularen Flüssigkeiten mit Dielectrischer Spektroskopie, Dissertation, Universität Mainz, *Verlag Shaker*, Aachen, **1995**.

55. Hecksher, T.; Nielsen, A.I.; Olsen, N.B.; Dyre, J.C. Little evidence for dynamic divergence in ultraviscous molecular liquids; Nat. Phys. **2008**, *4*, 737-741.

56. Mauro, J.C.; Yue, Y.; Ellison, A.J.; Gupta, P.K.; Allan, D.C. Viscosity of glass-forming liquids; *Proc. Natl. Acad. Sci. USA* **2009**, *106*, 19780.

57. Lunkenheimer, P.; Kastner, S.; Köhler, M.; Loidl, A. Temperature development of glassy α-relaxation dynamics determined by broadband dielectric spectroscopy, *Phys. Rev. E* **2010**, *81*, 051504.

58. Gainaru, C.; Lips, O.; Troshagina, A.; Kahlau, R.; Brodin, A.; Fujara, F.; Rössler, E.A. On the nature of the high-frequency relaxation in a molecular glass former: A joint study of glycerol by field cycling NMR, dielectric spectroscopy, and light scattering, *J. Chem. Phys.* **2008**, *128*, 174505.

59. Böhmer, R; Hinze, G., Reorientations in supercooled glycerol studied by two-dimensional time-domain deuteron nuclear magnetic resonance spectroscopy, *J. Chem. Phys* **1998**, *109*, 241-248.

60. Böhmer, R; Diezemann, G.; Hinze, G.; Rössler, E., Dynamics of supercooled liquids and glassy solids, *Prog. Nucl. Magn. Reson. Spectroscopy* **2001**, *39*, 191-267.

61. Richert, R. On the dielectric susceptibility spectra of supercooled *o*-terphenyl, *J. Chem. Phys.* **2005**, *123*, 154502.

62. Ruocco, G.; Sciortino, F.; Zamponi, F.; De Michele, C.; Scopigno, T. Landscapes and fragilities, *J. Chem. Phys.* **2004**, *120*, 10666.

63. McKenna, G.B. Looking at the glass transition: challenges of extreme time scales and other interesting problems, *Rubber Chemistry and Technology* **2020**, *93*, 79-120.

64. Wang, L.M. Enthalpy relaxation upon glass transition and kinetic fragility of molecular liquids, *J. Phys. Chem.* **2009**, *113*, 5168-5171.

65. Cohen, M.H.; Turnbull, D. Molecular transport in liquids and glasses, *J. Phys. Chem.* **1959**, *31*, 1164-1169.

66. Cohen, M.H.; Grest, G.S. Liquid-glass transition, a free-volume approach, *Phys. Rev. B* **1979**, *20*, 1077-1098.

67. Glarum, S.H.; Dielectric relaxation of isoamyl bromide, *J. Phys. Chem.* **1960**, *33*, 639-643.

68. Bendler, J.T.; Fontanella, J.J.; Shlesinger, M.F.; Wintersgill, M.C. The defect diffusion model, glass transition and the properties of glass-forming liquids, **2008**, *CP982, Complex Systems, 5th International Workshop on Complex Systems,* ed. Tokoyama, M.; Oppenheim, I.; Nishiyama, AIP Conferences.

69. Dyre, J.C.; Olsen, N.B.; Christensen, T. Local elastic expansion model for viscous-flow activation energies of glass-forming molecular liquids, *Phys. Rev. B* **1996**, *53*, 2171-2174.

70. Dyre, J.C. Colloquium: The glass transition and elastic models of glass-forming liquids, *Rev. Mod. Phys.* **2006**, *78*, 953-972.

71. Adam, G.; Gibbs, J.H. On the temperature dependence of cooperative relaxation properties in glass-forming liquids, *J. Phys. Chem.* **1965**, *43*, 139-146.

72. Yamamuro, O.; Tsukushi, I.; Lindqvist, A.; Takahara, S.; Ishikawa, M.; Matsuo, T. Calorimetric study of glassy and liquid toluene and ethylbenzene: Thermodynamic approach to spatial heterogeneity in glass-forming molecular liquids, *J. Phys. Chem. B* **1998**, *102*, 1605-1609.

73. Dyre, J.C. A brief critique of the Adam-Gibbs entropy model, *J. Non-Cryst. Solids* **2009**, *355*, 624-627.

74. Götze, W. Recent tests of the mode-coupling theory for glassy dynamics, *J. Phys.: Condens. Matter* **1999**, *11*, A1-A45.

75. Das, S.P. Mode-coupling theory and the glass transition in supercooled liquids, *Rev. Mod. Phys.* **2004**, *76*, 785-851.

76. Lubchenko, V.; Wolynes, P.G. Theory of structural glasses and supercooled liquids, *Annu. Rev. Phys. Chem.* **2007**, *58*, 235-266.

77. Chamberlin, R.V. Mean-field cluster model for the critical behaviour of ferromagnets, *Nature* **2000**, *408*, 337-339.

78. Moore, M.A. Interface free energies in *p*-spin glass models *Phys. Rev. Lett.* **2006**, *96*, 137202.

79. Yeo, J,; Moore, M.A. Possible instability of one-step replica symmetry breaking in *p*-spin Ising models outside mean-field theory, *Phys. Rev. E* **2020**, *101*, 032127.

80. Kivelson, D.; Kivelson, S.A.; Zhao, X.; Nussinov, Z.; Tarjus, G. A thermodynamic theory of supercooled liquids, *Physica A* **1995**, *219*, 27-38.

81. Tarjus, G.; Kivelson, S.A.; Nussinov, Z.; Viot, P. The frustration-based approach of supercooled liquids and the glass transition: a review and critical assessment, *J. Phys.: Condens. Matter* **2005**, *17*, R1143-R1182.

82. Tarjus, G. An overview of the theories of the glass transition, Chapt. 2 in *Dynamical heterogeneities in glasses, colloids, and granular media*, Eds.: Berthier, L.; Biroli, G.; Bouchaud, J.P.; Cipelleti, L.; van Saarloos, W.; Oxford University Press, Oxford, UK **2011**, pp. 39-67.

83. Chamberlin, R.V. Monte Carlo simulations including energy from an entropic constraint, *Physica A* **2012**, *391*, 5384-5391.